\documentclass[aps,pra,epsfigure,twocolumn,showpacs]{revtex4}
\usepackage{dcolumn}    
\usepackage{bm} 
\usepackage{graphicx}
\usepackage{amsmath}    
\usepackage{latexsym}
\usepackage{amsfonts}   
\usepackage{amssymb}
\usepackage{array}      
\usepackage{epsfig}
\usepackage{epstopdf}
\newcommand{\ket}[1]{\left\vert#1\right\rangle}

\newcommand{\ketbra}[2]{\left\vert#1\rangle\langle#2\right\vert}

\newcommand{\sprod}[2]{\left\langle#1\vert#2\right\rangle}

\newcommand{\bra}[1]{\left\langle#1\right\vert}

\newcommand\be{\begin{equation}}
\newcommand\ee{\end{equation}}
\newcommand\bea{\begin{eqnarray}}
\newcommand\eea{\end{eqnarray}}
\newcommand\beas{\begin{eqnarray*}}
\newcommand\eeas{\end{eqnarray*}}
\def\l({\left(}
\def\r){\right)}
\def\l[{\left[}
\def\r]{\right]}

\def\ba{\begin{array}}
\def\ea{\end{array}}

\newcommand{\KS}{_{\hbox{\footnotesize\sc ks}}}
\newcommand{\CHSH}{_{\hbox{\footnotesize\sc chsh}}}
\begin{document}
\title{Two proposals for testing quantum contextuality of continuous-variable states}
\author{Gerard McKeown$^1$, Matteo G. A. Paris$^2$, and Mauro Paternostro$^1$} 
\affiliation{$^1$School of Mathematics and Physics, Queen's University, Belfast
BT7 1NN, United Kingdom\\
$^{2}$Dipartimento di Fisica, Universita' degli Studi di Milano, I-20133 Milano, Italy\\
$^3$CNISM, Udr Milano, I-20133 Milan, Italy}
\date{\today}
\begin{abstract}
We investigate the violation of non-contextuality by a class of
continuous variable {states}, including variations of entangled coherent
states (ECS's) and a two-mode continuous superposition of coherent states. We generalise the
Kochen-Specker (KS) inequality discussed in A. Cabello, Phys. Rev. Lett.
{\bf 101}, 210401 (2008) by using effective bidimensional observables
implemented through physical operations acting on continuous variable
states, in a way similar to an approach to the falsification of
Bell-CHSH inequalities put forward recently. We test for
state-independent violation of KS inequalities under variable degrees of
state entanglement and mixedness. We then demonstrate theoretically the
violation of a KS inequality for any two-mode state by using
pseudo-spin observables and a generalized quasi-probability function.
\end{abstract}
%%%%
\maketitle
%%%%
\section{Introduction}
Non-contextuality is commonly intended as a property of mutually
compatible observables. Two observables $A$ and $B$ are said to be
compatible when  the outcome of a measurement of $A$ performed on a
system does not depend on any prior or simultaneous measurement of $B$.
A set of mutually compatible observables defines a context, so that the
above examples defines a situation where the measurement of A does not
depend on the context or is {\it non-contextual}. Clearly,
non-contextuality is a property inherent in the classical world. In a
maieutic game played by Alice and Bob, if Alice asks a question, then
clearly the answer is not affected by any prior or simultaneous
compatible question asked by Bob. 

For quantum observables to assume such a property may at first hand seem
reasonable. {It would be equally reasonable to assume functional
consistency (realism), i.e. for the commuting operators $A_1$, $A_2$ and
$A_3=f(A_1,A_2)$ to assume that the results of their measurements
(even if not performed) would satisfy the same relation as the
operators, e.g. $a_1$, $a_2$ and $f(a_1,a_2)$. On the other hand, the
two assumptions taken together are incompatible with quantum mechanics.}

In fact, the Kochen-Specker (KS) theorem~\cite{Specker1,
Specker2, Specker3} states that no non-contextual hidden variable (NCHV)
theory can reproduce quantum mechanics. This is complementary to the
well-known Bell theorem~\cite{bell}, which states that no local hidden
variable theory can reproduce quantum mechanics and provides an equally
viable tool to gaining insight into the open question to where exactly
the boundary between the classical and quantum may lie. Kochen and
Specker \cite{Specker2} originally produced a set of 117 observables,
associated with the squares of the components of the angular momentum
operator along 117 different directions to demonstrate a contradiction
with non-contextuality. Almost twenty five years later, Peres found a
much simpler counter-example~\cite{peres} involving only six Pauli spin
operators in the four-dimensional space of two spin-${1}/{2}$ particles.
Peres' formulation of the problem, however, is strictly dependent on the
form of the state of the two particles. Mermin~\cite{mermin} made a
further simplification by extending the example to include three
additional operators, thereby illustrating state-independence. The
state-independent nature of the KS theorem is a rather distinctive
feature: inequalities based on non-contextual hidden variable theories
(herein dubbed as KS inequalities) might be violated by any quantum
state, regardless of their degree of entanglement. 

It should be remarked how the falsification of a KS inequality faces
rather challenging hurdles related to the feasibility of tests that,
while are capable of maintaining state independence, also guarantee that
all the necessary  observables are measured in a context-independent
way~\cite{guhene}. Cabello~\cite{cabello} has recently addressed these
points by providing inequalities that strictly meet the criteria
mentioned above. One such inequality is built from the observables used
in the proof of the KS theorem for two qubit systems proposed by Peres
and Mermin \cite{peres,mermin}. In a seminal experiment, Kirchmair {\it
et al.}~\cite{Kirchmair} have demonstrated the violation of such an
inequality using trapped ions, were two energy levels of an ion are
selected so as to embody the single-qubit logical states. The KS
inequality was thus tested using ten different quantum states, ranging
from entangled to separable, from quasi-pure to almost fully mixed
states, hence providing compelling evidence of the state-independent
character of the inequality being probed. 

Any experimentally testable state-independent KS inequality proposed so
far deal with states belonging to Hilbert spaces of finite
dimension and dichotomic observables. Plastino and
Cabello~\cite{Plastino} have extended the notion of quantum
contextuality to include harmonic oscillators by deriving a KS
inequality involving 18 observables based on position and momentum.
Their conclusion is that it may indeed be possible to experimentally
reveal state-independent quantum contextuality for any quantum system
admitting two continuous position observables and corresponding
canonically conjugate momenta. However, the required measurements might
be quite demanding to implement in actual experiments using specific
physical systems.

Here, at variance with Plastino and Cabello, we tackle the falsification
of non-contextuality inequalities in unbounded Hilbert spaces using a
different viewpoint. In fact, while we keep the dichotomic structure of
the observable entering the KS inequalities to test, we explicitly
consider systems living in infinite-dimensional Hilbert spaces. In order
to accomplish our goal, we take advantage of the well-known possibility
to violate Bell-like inequalities using dichotomic non-Gaussian observables 
and continuous-variable (CV) systems prepared in quantum correlated Gaussian
states~\cite{Banaszek, Chen} as well as non-Gaussian states embedding a
qubit state~\cite{Jeong, Paternostro}. In particular, 
%One example of a CV system  possible for revealing a Bell violation
%using an extension to the usual qubit based Bell-CHSH type, is that of 
two-mode entangled coherent states (ECSs)~\cite{sanders} and binned
homodyne detections have been used by Stobi\'{n}ska {\it et
al.}~\cite{Stobinska} to show the violation of a
Bell-Clauser-Horn-Shimony-Holt (Bell-CHSH) inequality up to Tsirelson's
bound ({\it i.e.} the maximum degree of violation allowed by quantum
mechanics). In this case, the observables needed for the Bell-CHSH
inequality are given by effective rotations built form a series of Kerr
non-linearities, displacement operations and phase shifters. This
approach to `mimic' the standard Bell-CHSH inequality proved to be quite
efficient in demonstrating further the nonlocal properties of highly
mixed states close to classicality~\cite{jeong}, non-local
realism~\cite{MP,lee} and multipartite non-locality of a class of
multi-qubit states~\cite{GMcK}.

On a parallel line, motivated by feasibility in quantum optical systems,
dichotomic observables based on on/off photodetection have been extensively
employed to demonstrate violation of Bell-CHSH inequalities using realistic, 
not fully efficient, photodetectors with either qubit-like states or genuinely 
continuous-variable ones \cite{ban02,cinv05,so05}.  In this paper, we take a similar approach to show that a KS inequality
can be violated, in a state independent manner, using qubit states
encoded into genuinely infinite dimensional systems. We use the same
inequality as in Ref.~\cite{Kirchmair}, which is constructed by means of
the effective bidimensional observables that have been exploited for
Bell-CHSH inequalities mentioned above. While, on one hand, the number
of observables necessary for our task is strictly the same as for
discrete-variable systems, our proposal may pave the way to a
foreseeable experimental implementation faithful to the constraint of
context independence. We then further our study to test a KS inequality
using a class of states that do not embed an effective qubit state.This
makes the formulation of an analogy with the discrete-system case quite
problematic. The paradigm for such a situation is embodied by a two-mode
squeezed state. We overcome the difficulties by using the `pseudo-spin'
formalism introduced in Ref.~\cite{Chen}.  Maximum violation of the KS
inequality proves interesting for this class of states that, in the
limit of infinite squeezing, approximate the original version of the
Einstein-Podolski-Rosen (EPR) state and thus strengthen the claim on the
nonexistence of a hidden variable theory to describe quantum mechanics.
We generalize our approach by proving that it is indeed possible to
violate a KS inequality with any bipartite state of two harmonic
oscillators, such as two modes of the radiation field. To achieve this we used
a generalised $P$ representation to describe any two-mode
state~\cite{Glauber}. As we show, the application of  pseudo-spin
operators to construct the KS inequality warrants state-independence.

The remainder of this paper is organized as follows. In Sec.~\ref{KS} we
briefly review the KS formalism, introduce the non-contextual
inequality that will be tested throughout our work and introduce the
class of effective two-qubit operations with which we build up the
observables to be used. Sec.~\ref{WernerViolation} assesses the
violation of the KS inequality by a CV class of Werner state, which we
build using ECS form of entangled states (we defer to an Appendix the more formal aspects of our study). We show that, regardless of
the degree of entanglement and purity that characterize the states, a
large enough amplitude of the coherent states involved guarantees for
the state-independent maximum violation of a KS inequality. The case of
pseudo-spin operators applied to a test-bed state embodied by a two-mode
squeezed vacuum is discussed in Sec.~\ref{TMSV} and then generalized to
any two-mode state, towards full state independence, in
Sec.~\ref{quasiP}. Finally, in Sec.~\ref{conclusions} we summarize our
results and leave some open questions. 
%%%
\section{Kochen-Specker inequality and general formalism}
\label{KS} 
%%%
\subsection{The KS inequality}
We briefly introduce and discuss the KS inequality that has been
experimentally tested in Ref.~\cite{Kirchmair} and is assessed in this
paper. The inequality is constructed using nine observables, along the
lines of the arguments put forward by Peres and Mermin~\cite{peres,mermin} 
to prove the incompatibility between quantum mechanics and
non-contextuality. Such observables are arranged in a 3$\times$3 array
$\hat{\bm A}$, known as the Peres-Mermin square, in such a way that  the
entries $\hat A_{ij}~(i,j{=}1,2,3)$ in each column and row are mutually
compatible and  have dichotomic outcomes $\nu{(\hat A_{ij})}{=}\pm{1}$.
Denote the products of rows and columns as 
\begin{align*}
\hat{R}_k &= \hat{A}_{k1}\hat{A}_{k2}\hat{A}_{k3}
\\ 
\hat{C}_k &= \hat{A}_{1k}\hat{A}_{2k}\hat{A}_{3k}\,,
\end{align*}
respectively: Assuming non-contextuality implies that 
\begin{align*}
\nu(\hat{R}_k)&=\nu(\hat{A}_{k1})\nu(\hat{A}_{k2})\nu(\hat{A}_{k3})
\\
\nu(\hat{C}_k)&=\nu(\hat{A}_{1k})\nu(\hat{A}_{2k})\nu(\hat{A}_{3k})\,.
\end{align*}
Thus the total product becomes $\Pi_{k=1}^3\nu(R_k)\nu(C_k)=1$, 
since any $\nu(\hat{A}_{ij})$ appears twice in the product.
However, this is in contrast with the predictions of quantum
mechanics, where a Peres-Mermin square can be built out of the
dichotomic Pauli operators 
$$\hat\sigma_{x}=\begin{pmatrix}0&1\\
1&0\end{pmatrix},~~\hat\sigma_{y}=\begin{pmatrix}0&-i\\
i&0\end{pmatrix},~~\hat\sigma_{z}=\begin{pmatrix}1&0\\
0&-1\end{pmatrix},~~$$ associated with two spin-$1/2$ systems as
\begin{equation}
\hat{\bm A} =\left[
\begin{matrix}
\hat\sigma_{z}^{(1)}{\otimes}\hat\openone^{(2)}
&\hat\openone^{(1)}{\otimes}\hat\sigma_{z}^{(2)}
&
\hat\sigma_{z}^{(1)}{\otimes}\hat\sigma_{z}^{(2)}
\\
\hat\openone^{(1)}{\otimes}\hat\sigma_{x}^{(2)}&
\hat\sigma_{x}^{(1)}{\otimes}\hat\openone^{(2)}
&
\hat\sigma_{x}^{(1)}{\otimes}\hat\sigma_{x}^{(2)}
\\
\hat\sigma_{z}^{(1)}{\otimes}\hat\sigma_{x}^{(2)}
&\hat\sigma_{x}^{(1)}{\otimes}\hat\sigma_{z}^{(2)}
&\hat\sigma_{y}^{(1)}{\otimes}\hat\sigma_{y}^{(2)}
\end{matrix}
\right]
\label{MPer}
\end{equation}
%
%\begin{equation}	\begin{array}{clcr}
%			\hat A_{11}=\hat\sigma_{z}^{(1)}\otimes\hat\openon\text{e}^{(2)} &\hat A_{12}=\hat\sigma_{z}^{(1)}\otimes\hat\openon\text{e}^{(2)}  & \hat A_{13}=\hat\sigma_{z}^{(1)}~\otimes\hat\sigma_{z}^{(2)}   \\
%			A_{21}=\hat\sigma_{x}^{(1)}\otimes\hat\openon\text{e}^{(2)}    &\hat A_{22}=\hat\sigma_{x}^{(1)}\otimes\hat\openon\text{e}^{(2)}   & \hat A_{23}=\hat\sigma_{x}^{(1)}~\otimes\hat\sigma_{x}^{(2)}    \\
%			\hat A_{31}=\hat\sigma_{z}^{(1)}\otimes\hat\sigma_{x}^{(2)}      &\hat A_{32}=\hat\sigma_{x}^{(1)}\otimes\hat\sigma_{z}^{(2)} &\hat A_{33}=\hat\sigma_{y}^{(1)}\otimes\hat\sigma_{y}^{(2)}
%		\end{array}
%		\label{MPer}
%	\end{equation}
%where each entry $\hat A_{ij}~(i,j{=}1,2,3)$ has dichotomic outcomes given by $\nu{(\hat A_{ij})}{=}\pm1$.  Any two observables that share a subscript commute, hence observables positioned along any row or column are mutually compatible. 
In this case, the product of each row and column gives $\openone$, except those of the last column that gives $-\openone$. Hence, in this case we have the additional property of compatibility for $\hat{R}_k$ and $\hat{C}_k$ (k=1,2,3) and so, assuming non-contextuality, $\Pi_{k=1}^3\nu(\hat{R}_k)\nu(\hat{C}_k){=}\nu\left(\Pi_{k=1}^3\hat{R}_k\hat{C}_k\right){=}{-}1$.
This witnesses the contradiction between a
non-contextual assumption and the predictions of quantum mechanics. Such
a conflicting outcome is  formalized by the KS-like
inequality~\cite{cabello} 
\be
\label{KSineq} \langle\hat \chi_{\KS}\rangle=\langle \hat
R_1\rangle+\langle \hat  R_2\rangle+\langle \hat  R_3\rangle+\langle
\hat  C_1\rangle+\langle \hat  C_2\rangle-\langle \hat  C_3\rangle\leq4.
\ee
%where $\hat R_{k}=\hat A_{k1}\otimes\hat A_{k2}\otimes\hat A_{k3}$ and
%$\hat C_{k}=\hat A_{1k}\otimes\hat A_{2k}\otimes\hat A_{3k}$ for
%$k=1,2,3$.
In Ref.~\cite{cabello} it has been proven that this
inequality is bounded by 4 for any NCHV theory, while $\langle\hat
\chi_{\KS}\rangle=6$ for any state of two spin-$1/2$ particles.
Eq.~(\ref{KSineq}) will be used throughout this paper.
%%% 
\subsection{General formalism for CV states and effective bidimentional
dichotomic observables}
In this Section we introduce the class of CV states of interest and the
effective observables necessary for the falsification of the KS
inequality discussed above. The class of state that will be used in the
first part of our work are built on coherent states $\ket{\alpha}$
($\alpha{\in}\mathbb{C}$), which are obtained by applying the
displacement operator $\hat D(\alpha){=}\exp(\alpha \hat
a^\dagger{-}\alpha^{\ast} \hat a)$ to the vacuum state
$\ket{0}$~\cite{barnett}. Here, $\hat{a}$ ($\hat a^\dag$) is the annihilation 
(creation) operator of a bosonic system.
%, that is, $\ket{\alpha}=\hat D(\alpha)\ket{0}$. 
Although two coherent states with opposite phases $\ket{\alpha}$ and
$\ket{-\alpha}$ are strictly non-orthogonal, we have
$\sprod{\alpha}{-\alpha}{=}\exp[-2|\alpha|^2]{\rightarrow}0$ in the
limit of $\alpha{\gg}{1}$. In such conditions,
$\{\ket{\alpha},\ket{-\alpha}\}$ form a basis in a two-dimensional
Hilbert space. This reasoning paves the way for extending the KS
inequality in Eq.~(\ref{KSineq})  to deal with CV systems represented in
the coherent-state qubit basis. As it will be clarified in
Sec.~\ref{WernerViolation}, in our investigation we consider mixed
states of two coherent-state qubits having a variable degree of
entanglement between them. 

The second important point is embodied by the provision of appropriate
observables able to mimic the Pauli spin-${1}/{2}$ ones entering
$\hat{\bm A}$ in Eq.~(\ref{MPer}). To do this, we take advantage of the
results reported in~\cite{Stobinska,MP}, where effective rotations are
introduced in order to run a Bell-CHSH test. Such operations are
generally given by the 2$\times$2 transformation matrix acting on the
space spanned by the coherent-state qubit
$\{\ket{\alpha},\ket{-\alpha}\}$ 
\begin{equation}
 \label{rot}
 \hat{O}(\theta,\phi)=\left( \begin{array}{cc}
\sin \frac{\theta}{2} & \text{e}^{i\phi}\cos\frac{\theta}{2} \\
\text{e}^{-i\phi}\cos\frac{\theta}{2} & -\sin\frac{\theta}{2}\\ 
 \end{array} \right).
\end{equation}
For proper choices of parameters $\theta\in[0,\pi]$ and
$\phi\in[0,2\pi]$, any spin-$1/2$ transformation can be realized.
Eq.~(\ref{rot}) can be simulated by a sequence of building-block
operations given by displacement operations given by
$\hat{D}(i\eta/2\alpha)$ (for proper choices of $\eta\in\mathbb{C}$) and
the single-mode Kerr-like nonlinearity
$\hat{U}_{NL}{=}\exp[-i\pi(\hat{a}^\dagger\hat{a})^2/2]$~\cite{walls}.
More precisely, a simulation of Eq.~(\ref{rot}) is  provided by
\begin{equation}
\label{kerr}
\hat{O}(\theta,\phi)\simeq\hat{D}(-i\phi/4\alpha)\hat{U}_{NL}\hat{D}(i\theta/4\alpha)\hat{U}_{NL}\hat{D}(i\phi/4\alpha),
\end{equation}
where the symbol $\simeq$ is used to remind that the approximation improves as $|\alpha|$ grows.
The explicit transformations experienced by $\ket{\pm\alpha}$ are given by~\cite{MP}
\begin{equation}
\label{trasf}
\begin{aligned}
&\ket{\alpha}\rightarrow\frac{1}{2}\left[\text{e}^{\frac{i\theta}{4}}\left(\ket{\alpha+\frac{i\theta}{4\alpha}}+i\text{e}^{\frac{i\phi}{2}}\ket{-\alpha-\frac{i\phi}{2\alpha}-\frac{i\theta}{4\alpha}}\right)\right.\\
&\left. +i\text{e}^{-\frac{i\theta}{4}}\left(\text{e}^{\frac{i\phi}{2}}\ket{-\alpha-\frac{i\phi}{2\alpha}+\frac{i\theta}{4\alpha}}+i\ket{\alpha-\frac{i\theta}{4\alpha}}\right)\right],\\
&\ket{-\alpha}\rightarrow\frac{1}{2}\left[i\text{e}^{\frac{i\theta}{4}}\left(i\ket{-\alpha-\frac{i\theta}{4\alpha}}+\text{e}^{\frac{-i\phi}{2}}\ket{\alpha-\frac{i\phi}{2\alpha}+\frac{i\theta}{4\alpha}}\right)\right.\\
&\left. +\text{e}^{-\frac{i\theta}{4}}\left(i\text{e}^{-\frac{i\phi}{2}}\ket{\alpha-\frac{i\phi}{2\alpha}-\frac{i\theta}{4\alpha}}+\ket{-\alpha+\frac{i\theta}{4\alpha}}\right)\right].
\end{aligned}
\end{equation}
Eqs.~(\ref{trasf}) are crucial in the construction of $\langle\hat \chi_{\KS}\rangle$.

\begin{figure*}[t]
\center{{\bf (a)}}\hskip5.5cm{{\bf (b)}}\hskip5.5cm{{\bf (c)}}
\psfig{figure=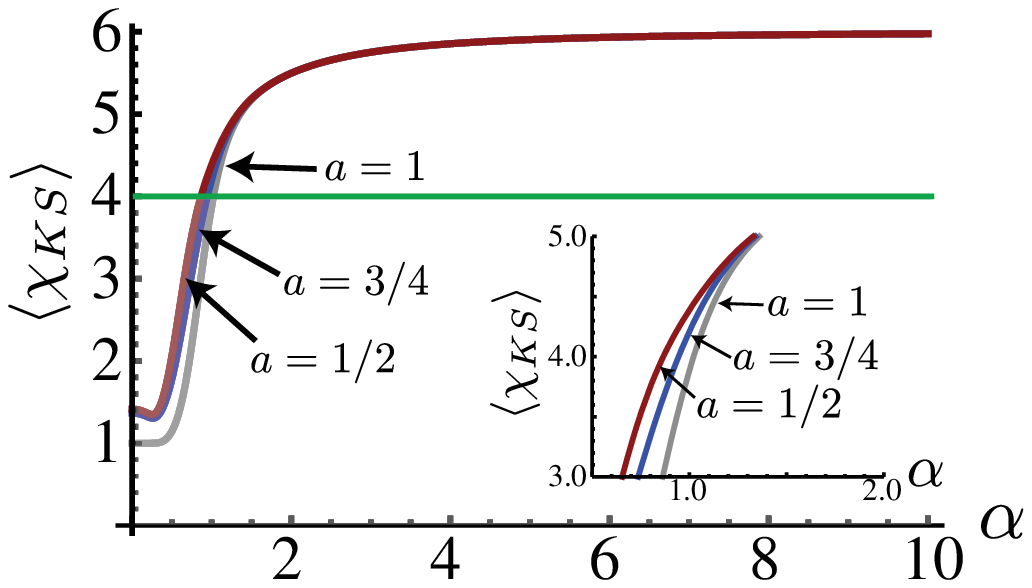,width=5.8cm,height=3.5cm}~~
\psfig{figure=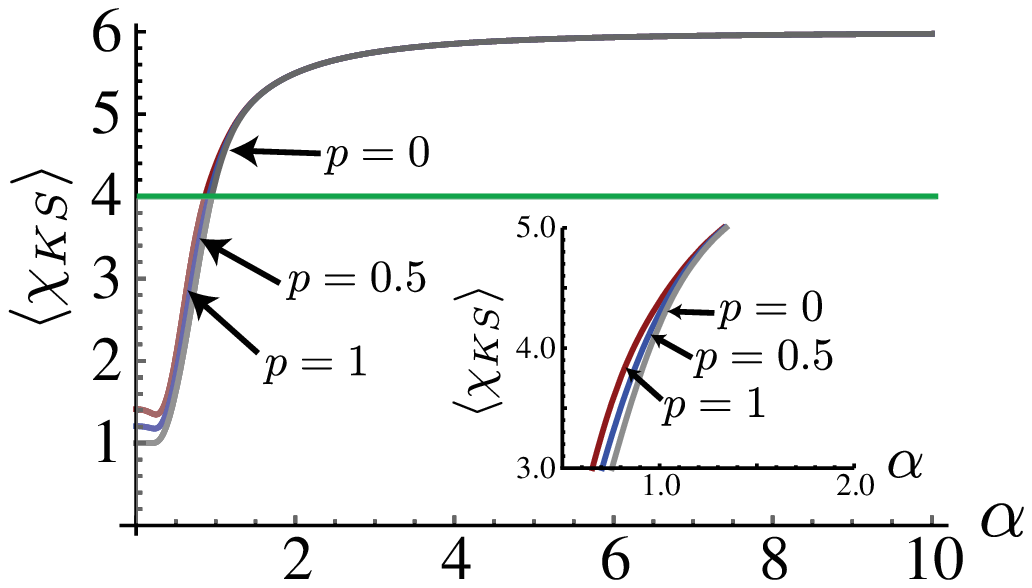,width=5.8cm,height=3.5cm}~~
\psfig{figure=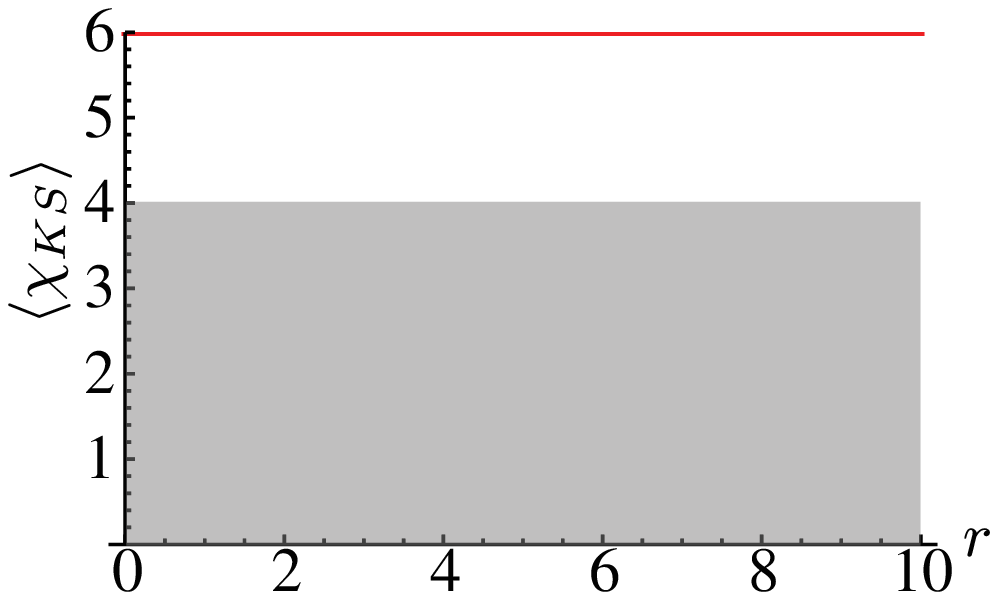,width=5.8cm,height=3.5cm}
\caption{{\bf (a)} Violation of non-contextuality by a CV Werner state
at increasing values of the amplitude $\alpha\in\mathbb{R}$. We plot  three KS
functions, each corresponding to $p=1$ in Eq.~(\ref{Werner}). The three
curves correspond to $a=1,3/4$ and $1/2$. Maximum violation of the
non-contextual KS inequality in Eq.~(\ref{KSineq}) is achieved
independently of the degree of entanglement. The inset shows a
magnification of the region given by $\alpha\in[0.5,2]$. In panel {\bf
(b)} we have plotted the KS functions corresponding to $a=0.5$, thereby
working with maximally entangled coherent states, for $p=1,0.5$ and $0$
in a CV Werner state. Maximum violation of the KS inequality is
achieved, regardless of the degree of mixedness within the state. The
inset shows a magnification of the region given by $\alpha\in[0.5,2]$.
{\bf (c)} Violation of KS inequality by a two mode-squeezed state
plotted against the squeezing parameter $r$. The KS inequality is given
by Eq.~(\ref{KSineq}), where the Pauli observables within each $\hat
A_{ij}$ are replaced by the corresponding pseudo-spin operator in
Eq.~(\ref{pseudospin}). In all the panels, the shaded region corresponds
to the constraints imposed by NCHV theories on the KS function.}
\label{wernerfig}
\end{figure*}
%%%%
\section{Violation of the KS inequality by a CV Werner state}
\label{WernerViolation}
Here, we discuss the performance of the KS inequality when tested using
a CV Werner-like class of states. These are defined as 
\begin{align}
\label{Werner}
\rho_w(a,p) =  & {p}\ketbra{\text{ECS}(a)}{\text{ECS}(a)} + \\ 
&\frac14 (1-p)\Big[\ketbra{\alpha,\alpha}{\alpha,\alpha}
 {+}\ketbra{\alpha,-\alpha}{\alpha,-\alpha}  \notag\\ 
& +\ketbra{-\alpha,\alpha}{-\alpha,\alpha}
  {+}\ketbra{-\alpha,-\alpha}{-\alpha,-\alpha}\Big]\,. \notag
\end{align}
State $\ket{\text{ECS}(a)}$ denotes a pure ECS reading 
$$\ket{\text{ECS}(a)}{=}N(\sqrt{a}\ket{\alpha,
\alpha}{+}\sqrt{1{-}a}\ket{-\alpha,-\alpha})\,,$$ 
whose degree of entanglement is  parameterised by $a\in[0,1]$
with $$N{=}[1{+}2\sqrt{(1-a)a}\text{e}^{-4|\alpha|^2}]^{-1/2}$$ being a
normalization factor. For $a{=}0,1$ the state is fully separable, while
at $a{=}{1}/{2}$ and $|\alpha|\gg{1}$ it approximates a maximally
entangled  two-qubit Bell state. The parameter $p\in[0,1]$ accounts for
the degree of mixedness of $\rho_w(a,p)$, which is a statistical mixture
(a pure ECS state) for $p=0$ ($p=1$). The combined tuning of $a$ and $p$
gives us access to a broad range of states that can be used to test the
KS inequality for a state-independent violation.

{The KS function $\langle\hat\chi_{\KS}\rangle$ for this Werner-like
class of states is built from the correlators $\langle \hat R_i\rangle$,
$\langle \hat C_i\rangle$  ($i=1,2,3$) as in Eq.~(\ref{KSineq}). Given
the general transformation matrix defined by Eq.~(\ref{rot}), the Pauli
spin-1/2 matrices $\hat{\sigma}_x$, $\hat{\sigma}_y$ and
$\hat{\sigma}_z$ are given by $\hat{O}(\theta=0,\phi=0)$,
$\hat{O}(\theta=0,\phi=-\pi/2)$ and $\hat{O}(\theta=\pi,\phi=0)$,
respectively. The correlator $\langle \hat \Gamma_i\rangle$ ($\hat\Gamma\in\{\hat R,\hat C\}$ and $i\in\{1,2,3\}$) is
written as 
\begin{equation}
\begin{aligned}
\langle \hat \Gamma_i\rangle&={(1-p)}\!\sum_{s_{1,2}=\pm}{\bra{s_1\alpha,s_2\alpha}\hat \Gamma_i\ket{s_1\alpha,s_2\alpha}}/{4}\\
&+p\bra{\text{ECS}(a)}\hat \Gamma_i\ket{\text{ECS}(a)}
%(\bra{\alpha,\alpha}\hat R_1\ket{\alpha,\alpha}+\bra{\alpha,-\alpha}\hat R_1\ket{\alpha,-\alpha}+\bra{-\alpha,\alpha}\hat R_1\ket{-\alpha,\alpha}+\bra{-\alpha,-\alpha}\hat R_1\ket{-\alpha,-\alpha})
%$,
\end{aligned}
\end{equation}
where
\begin{equation}
\begin{aligned}
&\bra{\text{ECS}(a)}\hat \Gamma_i\ket{\text{ECS}(a)}=a\bra{\alpha,\alpha}\hat \Gamma_i\ket{\alpha,\alpha}\\&+\sqrt{a(1-a)}(\bra{\alpha,\alpha}\hat \Gamma_i\ket{-\alpha,-\alpha}+h.c.)
%\\&\hspace{0.6cm}+\sqrt{a(1-a)}\bra{-\alpha,-\alpha}\hat R_1\ket{\alpha,\alpha}
\\&+(1-a)\bra{-\alpha,-\alpha}\hat \Gamma_i\ket{-\alpha,-\alpha}.
\end{aligned}
\end{equation}
Each correlator, $\hat\Gamma_i$, is given more explicitly in table 1 written in terms of the Pauli operators, described by the general transformation matrix in Eq.~(\ref{rot}).
%\begin{equation}
%\begin{aligned}
%$\hat R_1=[\hat O(\pi,0)^{(1)}\otimes\hat\openon\text{e}^{(2)}]\otimes[\hat\openon\text{e}^{(1)}\otimes\hat O(\pi,0)^{(2)}]\otimes[\hat O(\pi,0)^{(1)}\otimes\hat O(\pi,0)^{(2)}]$.
%\end{aligned}
%\end{equation}
\begin{table*}[ht]
\caption{Table providing the explicit products of general transformation matrices for each correlator building the KS function $\langle\hat\chi_{\KS}\rangle$.} % title of Table
\centering % used for centering table

\begin{tabular}{c l c c} % centered columns (4 columns)
\hline\hline %inserts double horizontal lines

$\hat\Gamma_i$ &~~~~~~~ Operator products entering the Peres-Mermin square\\
 % inserts table
%heading

\hline\hline \\[0.005ex] 
% inserts single horizontal line
$\hat R_1$ &~~~~~~~ $\left[\hat O^{(1)}(\pi,0)\otimes\hat\openone^{(2)}\right]\times\left[\hat\openone^{(1)}\otimes\hat O^{(2)}(\pi,0)\right]\times\left[\hat O^{(1)}(\pi,0)\otimes\hat O^{(2)}(\pi,0)\right]$  \\ [2.5 ex] % inserting body of the table
$\hat R_2$ &~~~~~~~ $\left[\hat\openone^{(1)}\otimes\hat O^{(2)}(0,0)\right]\times\left[\hat O^{(1)}(0,0)\otimes\hat\openone^{(2)}\right]\times\left[\hat O^{(1)}(0,0)\otimes\hat O^{(2)}(0,0)\right]$  \\[2.5ex] 
$\hat R_3$ &~~~~~~~ $\left[\hat O^{(1)}(\pi,0)\otimes\hat O^{(2)}(0,0)\right]\times\left[\hat O^{(1)}(0,0)\otimes\hat O^{(2)}(\pi,0)\right]\times\left[\hat O^{(1)}(0,-\pi/2)\otimes\hat O^{(2)}(0,-\pi/2)\right]$  \\[2.5 ex] 
$\hat C_1$ &~~~~~~~ $\left[\hat O^{(1)}(\pi,0)\otimes\hat\openone^{(2)}\right]\times\left[\hat\openone^{(1)}\otimes\hat O^{(2)}(0,0)\right]\times\left[\hat O^{(1)}(\pi,0)\otimes\hat O^{(2)}(0,0)\right]$  \\[2.5 ex] 
$\hat C_2$ &~~~~~~~ $\left[\hat\openone^{(1)}\otimes\hat O^{(2)}(\pi,0)\right]\times\left[\hat O^{(1)}(0,0)\otimes\hat\openone^{(2)}\right]\times\left[\hat O^{(1)}(0,0)\otimes\hat O^{(2)}(\pi,0)\right]$  \\[2.5ex] 
$\hat C_3$ &~~~~~~~ $\left[\hat O^{(1)}(\pi,0)\otimes\hat O^{(2)}(\pi,0)\right]\times\left[\hat O^{(1)}(0,0)\otimes\hat O^{(2)}(0,0)\right]\times\left[\hat O^{(1)}(0,-\pi/2)\otimes\hat O^{(2)}(0,-\pi/2)\right]$  \\[2.5ex]  % [1ex] adds vertical space
\hline %inserts single line
\end{tabular}
\label{table:nonlin} % is used to refer this table in the text
\end{table*}
The decomposition and effective realization of the transformation
matrices $\hat O(0,0)$, $\hat O(0,-\pi/2)$ and $\hat O(\pi,0)$ are given
in Eq.~(\ref{kerr}), while the result of their application to $\ket{\pm
\alpha}$ is determined by using Eq.~(\ref{trasf}). The corresponding
explicit form of the correlators as functions of $\alpha$, $p$ and $a$
are too lengthy to be shown here. However, the expressions are analytic
and allow us to obtain the full behavior of the KS function
$\langle\hat\chi_{\KS}\rangle$ against $\alpha$, for any degree of
entanglement and mixedness.}  
 
 \begin{figure}[b]
\hskip1.0cm\center{{\bf (a)}}\hskip4.2cm{{\bf (b)}}
\psfig{figure=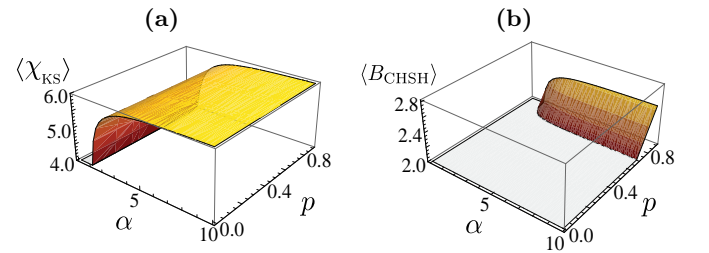,width=8.8cm,height=3.0cm}
\caption{(Color online) {\bf (a)}: Violation of the KS inequality Eq.~(\ref{KSineq}) by a CV Werner state with $a=1/2$. We plot $\langle\chi_{\KS}\rangle$ against the amplitude $\alpha$ and the purity parameter $p$. As discussed in Fig.~\ref{wernerfig} {\bf (b)}, the violation of the KS inequality is quasi-insensitive to variations of $p$. {\bf (b)}: Violation of Bell-CHSH inequality by a CV Werner state with $a=1/2$, plotted against $\alpha$ and $p$. As purity grows, $\langle\hat{B}_{\text{CHSH}}\rangle$ violates the local realistic bound of $2$.}
\label{confronto}
\end{figure}

In Fig.~\ref{wernerfig} {\bf (a)} and {\bf (b)}, we show two significant
cases of the quasi-state independence of the KS function
$\langle\hat\chi_{\KS}\rangle$ achieved in our model (for simplicity, we have taken $\alpha\in\mathbb{R}$). 
%Here we have two plots illustrating state independent maximal violation
%of the KS inequality across a range of states for specific values of
%$a$ and $p$. 
{The analytic expression for each KS function is given, for completeness, in the Appendix.} Here, we focus on the general features of such functions. Panel {\bf (a)} is for $p=1$ and three different values of the
entanglement within the state, from full separability to maximum
entanglement. On the other hand, panel {\bf (b)} studies the effects
that mixedness has on the behavior of $\langle\hat\chi_{\KS}\rangle$. We
set $a=0.5$, so that the CV Werner state is maximally entangled, and
tune $p$ from a fully pure state to maximum mixedness. The results are
clear: at small amplitudes of the coherent states involved in
$\rho_{w}(a,p)$, the KS function is an increasing function that
trespasses  the bound imposed by NCHV's in a narrow region around
$\alpha{\sim}1$. In these conditions, we observe some minor dependence
of the KS function from the various states being used. Those having
larger degrees of entanglement and purity become larger than $4$ for
slightly smaller values of $\alpha$. The situation changes as the
amplitude grows, nullifying the differences highlighted above and
delivering a truly state-independent KS function that quickly reaches
$6$, the value that is known to be achieved by
$\langle\hat\chi_{\KS}\rangle$ in the discrete-variable case and
regardless of the state being used. Although Figs.~\ref{wernerfig} {\bf
(a)} and {\bf (b)} address only a few significant cases, we have checked
that the description provided here is valid for any other choice of $a$
and $p$.

It is also interesting to compare the predictions for non-classicality 
given by the violation of a KS inequality to those regarding the violation 
of a local realism~\cite{Specker3,bell}. By following the approach described and used in Refs.~\cite{Stobinska,jeong,MP}, one can easily build up the Bell-CHSH function $\hat B_{\CHSH}$ associated with state $\rho_{w}(a,p)$ in Eq.~(\ref{Werner}) by means of local rotations realized through the operator $\hat O(\theta,\phi)$ and dichotomized homodyne projections onto quadrature eigenstates. In the qubit case, the violation the Bell-CHSH inequality requires rotations performed only on the equatorial plane of the Bloch sphere and this feature is carried over to the case at hand here. We thus have to consider the set of transformations (\ref{trasf}) obtained by setting $\phi{=}0$. Moreover, we can restrict the study to projections onto the eigenstates $\ket{x}$ of the position-like quadrature $(\hat{a}+\hat{a}^\dag)/\sqrt 2$ of a bosonic system~\cite{barnett}. Using the general formula for the projection of a coherent state $\ket\alpha$ (with $\alpha=\alpha_r+i\alpha_i$) over a position-like quadrature eigenstates $\ket{x}$
\begin{equation}
\langle{x}|\alpha\rangle=\frac{1}{\sqrt[4]{2\pi}}\text{e}^{i\alpha\alpha_i-(x/2-\alpha)^2}
\end{equation}
%%%
one can evaluate $\hat{B}_{\CHSH}$ analytically and then maximize it numerically over the parameters of the local rotations. The results are given in Fig.~\ref{confronto}, where a comparison is performed between the Bell-CHSH function and the KS one corresponding to $\rho_{w}(1/2,p)$ (we have taken $a=1/2$ here simply as a significant representative of the general behavior observed for an arbitrary choice of $a$). While panel {\bf (a)} summarizes the findings reported in Fig.~\ref{wernerfig} {\bf (b)}, {\it i.e.} the quasi-independence of $\langle\hat\chi_{\KS}\rangle$ of the value of $p$ entering the state under scrutiny, panel {\bf (b)} shows the sensitivity of a Bell-CHSH test to the degree of mixedness of $\rho_{w}(1/2,p)$. This is in line with the idea that KS tests are expected to be generally more powerful than CHSH ones 
in revealing the quantumness of a physical system.

\section{Violation of the KS inequality by a continuous superposition of coherent states}
\label{TMSV}

We would like now to extend the class of systems that we use for our
goals from the discrete superposition of quasi-orthogonal states that
builds up an ECS to a continuous distribution.  As the archetypal
example of such case, we consider the state produced by
superimposing a single-mode squeezed state to a vacuum mode at a 50:50
beam splitter~\cite{barnett,knight}. The former can be written in the
coherent-state basis as the continuous Gaussian-weighted distribution
${\cal N}\int\!d\alpha\,{\cal G}(r,\alpha)\ket\alpha$ with
\begin{equation}
{\cal
G}(r,\alpha)={\text e}^{-\frac{(1-\tanh r)\alpha^2}{2\tanh r}},~~~(\alpha\in\mathbb{R})
\end{equation}
where $r$ is the squeezing parameter and ${\cal N}{=}1/\sqrt{2 \pi
\sinh r}$ is the normalization factor~\cite{knight}. It is worth stressing that such a choice of resource state does not limit the validity of the results to come and is merely due to the experimental-friendly nature of the state, which can be routinely produced in many linear-optics labs. Any other choice would be equally valid for our purposes. 
After the admixture at the beam splitter, we get the two-mode
state~\cite{Paternostro}   
\begin{equation}
\label{sq}
\ket{\xi}={\cal N} \int\!\! d\alpha\,{\cal G}(r,\alpha)\,
|{{\alpha}/{\sqrt{2}},{\alpha}/{\sqrt{2}}}\rangle.
\end{equation} 
For this class of states any attempt to violate the KS inequality given
in Eq.~(\ref{KSineq}) by applying the same set of observables as done
for the Werner state, would be meaningless because of the difficulties
in identifying, in $\ket\xi$, a bipartite bidimensional system: although
the series of displacement operators and Kerr-like nonlinearities
introduced above can be used to sufficiently approximate the Pauli
matrices entering each $\hat A_{ij}$, the possibility of relating the
state $\ket\xi$ to that of a `two-qubit coherent state' is
undermined by the continuous nature of the distribution in
Eq.~(\ref{sq}). 

However it is not futile to try and falsificate the KS inequality
(\ref{KSineq}) by choosing a different set of observables than those
assessed so far. Our reasoning originates from the results by Chen {\it
et al.}~\cite{Chen}, whereby a generalisation of the Bell-CHSH
inequality for two-qubit systems to a two-mode state obtained superimposing
  a single-mode squeezed vacuum state with a vacuum state at a
  balanced beam splitter has
been shown to be possible. In Ref.~\cite{Chen}, the Bell-CHSH function
for the two-mode squeezed vacuum state is built from ``pseudo-spin"
operators having the form 
\begin{equation}
\begin{aligned}
\label{pseudospin}
\hat s_x&=\sum_{n=0}^\infty\left(\ketbra{2n+1}{2n}+\ketbra{2n}{2n+1}\right),\\
\hat s_y&=i\sum_{n=0}^\infty\left(\ketbra{2n}{2n+1}-\ketbra{2n+1}{2n}\right),\\
\hat s_z&=\sum_{n=0}^\infty\left(\ketbra{2n+1}{2n+1}-\ketbra{2n}{2n}\right).
\end{aligned}
\end{equation}
Here, $\ket{n}$ is a Fock state of $n$ excitations. {These operators share identical commutation relations to those of the spin-1/2 systems and for this reason} the vector $\hat{\bm
s}{=}(\hat s_x,\hat s_y,\hat s_z)$ can be regarded as the counterpart of
the Pauli one $\hat{\bm \sigma}=(\hat \sigma_x,\hat \sigma_y,\hat
\sigma_z)$. It acts upon the parity space of a boson and is for this
reason dubbed as a vector of parity-spin operators. 
Pseudospin operators have been used to reveal bipartite and 
tripartite nonlocality for quantum states with positive Wigner function
\cite{Chen,Fer05}. It can be seen as
a generalization to continuous variable systems of the one
introduced by Gisin and Peres for the case of discrete variable
systems \cite{gis92}, hence, for the case of a pure bipartite system, it
is equivalent to an entanglement test \cite{Jeong}.
\par
Our goal here is to
prove that the KS function in Eq.~(\ref{KSineq}) can indeed be tested
using $\hat{\bm s}$ and $\ket\xi$. This is straightforwardly done by
replacing each Pauli spin operator $\hat \sigma_l~(l{=}x,y,z)$ present
in each $\hat A_{ij}$ with the analogous pseudo-spin operator $\hat
s_l$. Each $\hat R_k$ and $\hat C_k$ is then constructed in the same
fashion as in Sec.~\ref{KS}. The expectation value of the operator
$\hat \Gamma_i$ over the state $\ket\xi$ is given by 
\begin{equation}
\bra{\xi}\!\hat \Gamma_i\!\ket{\xi}{=}{\cal N}^2
\int
\!\!d\alpha\, d\beta\,
{\cal G}(r,\alpha){\cal G}(r,\beta)
\bra{{\alpha'},{\alpha'}}\hat\Gamma_i \ket{{\beta'},{\beta'}}\,,
\end{equation}
where $\gamma'{=}\gamma/\sqrt 2~(\gamma{=}\alpha,\beta)$. 

Given that
%\begin{equation}
%\begin{aligned}
$\hat{s}_x\hat{s}_x{=}\hat\openone, \hat{s}_z\hat{s}_z{=}\hat\openone,
\hat{s}_z\hat{s}_x\hat{s}_y{=}i\hat\openone, \hat{s}_x\hat{s}_z\hat{s}_y{=}-i\hat\openone$,
it easily follows that,
% the operators $\hat R_j~(j{=}1,2,3)$ and 
%$\hat C_k~(k{=}1,2)$ are equal to $\hat\openone$ while $\hat C_3=-\hat\openone$. Thus, 
for $\hat\Gamma_i=\hat R_1, \hat R_2, \hat R_3, \hat C_1, \hat C_2$, we have %the correlator $\langle\hat\Gamma_i\rangle$ is given by
\begin{equation}
\bra{\xi}\!\hat \Gamma_i\!\ket{\xi}{=}{\cal N}^2
\int
\!\!d\alpha\, d\beta\,
{\cal G}(r,\alpha){\cal G}(r,\beta)\left
\langle{{\alpha'}}|{\beta'}\right\rangle^2{=}1\,
\end{equation}
while
% $\langle\hat C_3\rangle$,
\begin{equation}
\bra{\xi}\!\hat C_3\!\ket{\xi}{=}{\cal N}^2
\int
\!\!d\alpha\, d\beta\,
{\cal G}(r,\alpha){\cal G}(r,\beta)i^2\left
\langle{{\alpha'}}|{\beta'}\right\rangle^2{=}-1.
\end{equation}
Fig.~\ref{wernerfig} {\bf (c)} shows the behavior of the resulting KS
function against the squeezing parameter $r$. Clearly, the KS inequality
is maximally violated for any degree of squeezing. This result is in
virtue of the perfect dichotomization of the unbound Hilbert space where
$\ket\xi$ lives performed by the parity-spin operators. This is in
contrast with what is obtained for the violation of the Bell-CHSH
inequality, which occurs only within a finite window of squeezing.
However, the difference stems from the explicit state-dependent nature
of the non-locality inequalities, which is in striking contrast with the
state independence typical of a KS inequality. Incidentally, we see that
the original EPR state, , which is the limiting case of $|\xi\rangle$
  for infinite squeezing, maximally violates the KS inequality.  
%%%v%
\section{State independence of the KS inequality with pseudo-spin operators}
\label{quasiP}
%%%v%
So far we have successfully shown the violation of the KS inequalities
for important classes of CV systems, including pure, mixed, entangled
and separable ones. Yet the range of states that have been used to probe
the KS inequality is still limited and a generalization able to
undeniably prove the claimed state independence will be highly
desirable. This is what we do in this Section, where state independence
is verified using the picture given by the generalized quasi-probability
function~\cite{Glauber} of a two-mode bosonic state and pseudo-spin
operators. 
%The required observables that are sufficient to construct
%such a KS function again take the form of the pseudo-spin operators in
%Eq.~(\ref{pseudospin}). Hence the KS inequality we will use for the
%purpose of illustrating state-independent violation takes the form of
%Eq.~(\ref{KSineq}) with each Pauli operator $\hat \sigma_l$ replaced
%with its analogous pseudo-spin operator $\hat s_l$. 

The density operator $\rho$ of a single-mode state is given by the
Glauber $R$-representation~\cite{Glauber}, which is based on a function
of two complex variables $R(\alpha^*,\beta)$, analytic throughout the
finite $\alpha^*$ and $\beta$ planes, and given by 
\begin{equation}
R(\alpha^*,\beta)=\bra{\alpha}\rho\ket{\beta}\exp[(|\alpha|^2+|\beta|^2)/2].
\end{equation}   
Given the knowledge of $R(\alpha^*,\beta)$, the density operator is then 
written as 
\begin{equation}
\rho=\frac{1}{\pi^2}\int
\!\!d^2\alpha\, d^2\beta\,
\ket{\alpha}R(\alpha^*,\beta)\bra{\beta}
\text{e}^{-\frac{1}{2}\left(|\alpha|^2+|\beta|^2\right)}
\end{equation}   
with the normalization condition
%\begin{align}
${\pi}^{-1}\int R\left(\beta^*,\beta\right)\text{e}^{-|\beta|^2}d^2 \beta{=}1$.
%\end{align}  
These expressions are easily generalized to the case of a two-mode state, where
\begin{equation}
\begin{aligned}
R(\alpha_1^*,\alpha_2^*,\beta_1,\beta_2)=&\bra{\alpha_1,\alpha_2}\rho\ket{\beta_1,\beta_2}\text{e}^{\frac{1}{2}\sum^2_{j=1}(|\alpha_j|^2+|\beta_j|^2)}
\end{aligned} 
\end{equation}
and the density operator
\begin{equation}
\begin{aligned}
\rho&=\frac{1}{\pi^2}\int\ket{\alpha_1,\alpha_2}R(\alpha_1^*,\alpha_2^*,\beta_1,\beta_2)\bra{\beta_1,\beta_2}\\
&\times\text{e}^{-\frac{1}{2}\left(||\alpha_1|^2+|\beta_1|^2+|\alpha_2|^2+|\beta_2|^2\right)}d^2\alpha_1 d^2\beta_1d^2\alpha_2 d^2\beta_2
\end{aligned} 
\end{equation}
with the normalization 
\begin{equation}
{\pi}^{-1}\int R\left(\beta_1^*,\beta_2^*,\beta_1,\beta_2\right)\text{e}^{-|\beta_1|^2-|\beta_2|^2}d^2
\beta_1d^2 \beta_2=1.
\end{equation}  
%The KS function is then given by 
As we have that 
\begin{equation}
  \langle\beta_2, \beta_1|\hat\Gamma_i|\alpha_1, \alpha_2\rangle =1
\end{equation}
for $\hat\Gamma_i = R_1, R_2, R_3, C_1, C_2$, while 
\begin{equation}
\langle\beta_2, \beta_1|C_3|\alpha_1, \alpha_2\rangle = -1,
\end{equation}
the KS function, which is given by
\begin{align}
\left\langle\hat\chi_{\KS}\right\rangle&=\text{Tr}\{\rho\hat R_1\}+\text{Tr}\{\rho\hat R_2\}+\text{Tr}\{\rho\hat R_3\}\nonumber\\&+\text{Tr}\{\rho\hat C_1\}+\text{Tr}\{\rho\hat C_2\}-\text{Tr}\{\rho\hat C_3\},
\end{align}  
equals $\langle\chi_KS\rangle=6$ for any $R$ function, that is without any limitation imposed on the details of the state used in order to calculate the KS function. We can thus conclude that NCHV models are falsified in a state independent manner, which proves our claim.  

\section{conclusions}
\label{conclusions}

We have proposed a means for the violation of the KS inequality by an
ample variety of two-mode CV states. The first class of states that we
have used in order to discuss this issue is based primarily on mixed
ECS's, mimicking the family of two-mode Werner states. For this case, we
have found that effective bidimensional observables achieved through a
sequence of displacements and non-linear interactions are well suited
for proving the quasi state-independent violation of the KS inequality.
The independence from the details of the state being used becomes
rigorous under the limit of large-amplitude coherent states, when the CV
Werner family mimicks in an excellent way the discrete-variable
counterpart.

We extended the study to include a more general class of states, using
as a prototypical example the two-mode state obtained by superimposing a single-mode squeezed state to a mode prepared in vacuum. In this case, the
use of pseudo-spin operators in place of the usual Pauli-spin operators
was proven adequate for the desired task: the violation of a KS
inequality was proven to be maximum and rigorously state-independent.
Such claim has been strengthened by relying on the Glauber $R$-representation of any two-mode bosonic system.

Our study should be regarded as an attempt to extend the domain of
applicability of already formalized frameworks for the violation of NCHV
theories in general CV states. Certainly, some open questions remain to
be addressed in a more extensive way, especially in relation to the
experimental feasibility of compatible measurements to be performed over
the test state. We are currently investigating this point and the
possibility of employing weak measurement for the effective
implementation of the required set of measurements in a non-intrusive
way~\cite{incorso}. We conclude our analysis by commenting on the
existence of at least an experimental setting where the ingredients
required by our protocol for the violation of the KS inequality by CV
Werner-like states are all present (at least one by one). In particular,
we can consider systems consisting of nano-mechanical oscillators
coupled to superconducting qubits operating in the charge regime, which
have been the center of an extensive experimental and theoretical
interest in the last ten years~\cite{blencowe}. While the oscillators
would embody the bosonic modes onto which we encode the state of our CV
system, the coupling with the superconducting qubit can be tuned so as
to effectively engineer an ECS state of the mechanical systems and
realize both the displacement operation and non-linearities of the
Kerr-like form, thus potentially providing the whole toolbox needed in
our proposal~\cite{engineering}. Alternatively, we can use coupled
superconducting co-planar resonators or a bimodal resonator with an
embedded charge qubit~\cite{jacobs}, which effectively mimick the same
sort of situation described above and have the potential to implement
the very same type of effective interactions. We are currently
investigating the feasibility of a proof-of-principle test to be
conducted along these lines~\cite{incorso}.
%%%
\acknowledgments 
GMcK thanks Mr. Ciaran McKeown for discussions. We thank the EPSRC
(EP/G004579/1) and the British Council/MIUR British-Italian Partnership
Programme 2009-2010 for financial support. 
%%%%
\renewcommand{\theequation}{A-\arabic{equation}}
\setcounter{equation}{0}
\section*{APPENDIX: Analytic expressions for the KS functions in Sec.~\ref{WernerViolation}}

In this Appendix we provide the explicit analytic expressions for the KS functions used in Sec.~\ref{WernerViolation}.
% Fig.~\ref{wernerfig} {\bf (a)} and {\bf (b)} the KS functions are given expilticly below.
 We distinguish each function by considering the explicit dependence of $\langle\hat\chi_{\KS}\rangle$ on parameters $p$ and $a$. That is, we consider $\langle\hat\chi_{\KS}\rangle=\langle\hat\chi_{\KS}(p,a)\rangle$. We have
\begin{widetext}
\begin{equation}
\begin{aligned}
   &\langle\hat\chi_{\KS}(1,1)\rangle=3\text{e}^{-\frac{1024 \alpha ^8+96 \pi ^2 \alpha ^4+\pi ^4}{256 \alpha ^6+16 \pi ^2 \alpha
   ^2}} \left[\text{e}^{4 \alpha ^2}+\text{e}^{\frac{32 \alpha ^6}{16 \alpha ^4+\pi ^2}} \sin \left(\frac{\pi
   ^3}{32 \alpha ^4+2 \pi ^2}\right)+\text{e}^{\frac{6 \pi ^2 \alpha ^2}{16 \alpha ^4+\pi
   ^2}}-\text{e}^{\frac{6 \pi ^2 \alpha ^2}{16 \alpha ^4+\pi ^2}} \sin \left(\frac{8 \pi  \alpha ^4}{16
   \alpha ^4+\pi ^2}\right)\right.\\&\left.+\text{e}^{\frac{2 \alpha ^2 \left(32 \alpha ^4+\pi ^2\right)}{16 \alpha
   ^4+\pi ^2}} \sin \left(\frac{8 \pi  \alpha ^4}{16 \alpha ^4+\pi ^2}\right)-\text{e}^{4 \alpha ^2}
   \sin \left(\frac{\pi ^3}{32 \alpha ^4+2 \pi ^2}\right)-\text{e}^{\frac{4 \alpha ^2 \left(8 \alpha
   ^4+\pi ^2\right)}{16 \alpha ^4+\pi ^2}} \sin \left(\frac{\pi ^3}{32 \alpha ^4+2 \pi
   ^2}\right)+2 \text{e}^{4 \alpha ^2} \cos ^2\left(\frac{4 \pi  \alpha ^4}{16 \alpha ^4+\pi
   ^2}\right)\right.\\&\left.+\text{e}^{\frac{64 \alpha ^6+6 \pi ^2 \alpha ^2}{16 \alpha ^4+\pi ^2}}+2 \text{e}^{\frac{64
   \alpha ^6+6 \pi ^2 \alpha ^2}{16 \alpha ^4+\pi ^2}} \sin \left(\frac{8 \pi  \alpha ^4}{16
   \alpha ^4+\pi ^2}\right)-2 \text{e}^{\frac{32 \alpha ^6+6 \pi ^2 \alpha ^2}{16 \alpha ^4+\pi ^2}}
   \sin \left(\frac{\pi ^3}{32 \alpha ^4+2 \pi ^2}\right)+2 \text{e}^{\frac{1024 \alpha ^8+96 \pi ^2
   \alpha ^4+\pi ^4}{256 \alpha ^6+16 \pi ^2 \alpha ^2}}\right.\\&\left.-4 \text{e}^{\frac{1024 \alpha ^8+192 \pi ^2
   \alpha ^4+\pi ^4}{512 \alpha ^6+32 \pi ^2 \alpha ^2}} \sin \left(\frac{\pi ^3}{64 \alpha ^4+4
   \pi ^2}\right)+4 \text{e}^{\frac{2048 \alpha ^8+192 \pi ^2 \alpha ^4+\pi ^4}{512 \alpha ^6+32 \pi ^2
   \alpha ^2}} \cos \left(\frac{\pi ^3}{64 \alpha ^4+4 \pi ^2}\right)\right]
\end{aligned}   
\end{equation}

\begin{equation}
\begin{aligned}
&\langle\hat\chi_{\KS}(1,3/4)\rangle=\frac{\text{e}^{-\frac{768 \alpha ^8+96 \pi ^2 \alpha ^4+\pi ^4}{128 \alpha ^6+8 \pi ^2 \alpha ^2}}}{4 \sqrt{\sqrt{3} \text{e}^{-4 \alpha ^2}+2}}
   \left\{4 \left(-\text{e}^{2 \alpha ^2}+2 \text{e}^{\frac{2 \alpha ^2 \left(32 \alpha ^4+\pi ^2\right)}{16
   \alpha ^4+\pi ^2}}+\sqrt{3}\right) \text{e}^{\frac{1024 \alpha ^8+320 \pi ^2 \alpha ^4+3 \pi
   ^4}{512 \alpha ^6+32 \pi ^2 \alpha ^2}} \cos \left(\frac{4 \pi  \alpha ^4}{16 \alpha
   ^4+\pi ^2}\right)\right.\\&\left.+\text{e}^{\frac{512 \alpha ^8+\pi ^4}{256 \alpha ^6+16 \pi ^2 \alpha ^2}} \left[4
   \text{e}^{\frac{320 \pi ^2 \alpha ^4+\pi ^4}{512 \alpha ^6+32 \pi ^2 \alpha ^2}} \left(\text{e}^{2 \alpha
   ^2}+2 \text{e}^{\frac{2 \alpha ^2 \left(32 \alpha ^4+\pi ^2\right)}{16 \alpha ^4+\pi
   ^2}}+\sqrt{3}\right) \sin \left(\frac{4 \pi  \alpha ^4}{16 \alpha ^4+\pi ^2}\right)+\sqrt{2}
   \left(\left(2 \left(\sqrt{3}-1\right) \text{e}^{\frac{12 \pi ^2 \alpha ^2}{16 \alpha ^4+\pi ^2}}\right.\right.\right.\right.\\&\left.\left.\left.\left.+2
   \text{e}^{\frac{8 \alpha ^2 \left(8 \alpha ^4+\pi ^2\right)}{16 \alpha ^4+\pi
   ^2}}-\left(\sqrt{3}-4\right) \text{e}^{\frac{64 \alpha ^6+12 \pi ^2 \alpha ^2}{16 \alpha ^4+\pi
   ^2}}+\sqrt{3}\right) \sin \left(\frac{8 \pi  \alpha ^4}{16 \alpha ^4+\pi
   ^2}\right)-\text{e}^{\frac{32 \alpha ^6+6 \pi ^2 \alpha ^2}{16 \alpha ^4+\pi ^2}} \left(\text{e}^{\frac{4
   \pi ^2 \alpha ^2}{16 \alpha ^4+\pi ^2}}+2 \text{e}^{\frac{6 \pi ^2 \alpha ^2}{16 \alpha ^4+\pi
   ^2}}-1\right)\right.\right.\right.\\&\left.\left.\left.\times \sin \left(\frac{\pi ^3}{32 \alpha ^4+2 \pi ^2}\right)+\text{e}^{\frac{6 \pi ^2 \alpha
   ^2}{16 \alpha ^4+\pi ^2}} \left(4 \text{e}^{4 \alpha ^2}+\left(2+\sqrt{3}\right) \text{e}^{\frac{6 \pi ^2
   \alpha ^2}{16 \alpha ^4+\pi ^2}}+\left(2+\sqrt{3}\right) \text{e}^{\frac{64 \alpha ^6+6 \pi ^2 \alpha
   ^2}{16 \alpha ^4+\pi ^2}}+2 \sqrt{3} \text{e}^{\frac{32 \pi ^2 \alpha ^4+\pi ^4}{256 \alpha ^6+16 \pi
   ^2 \alpha ^2}}\right.\right.\right.\right.\\&\left.\left.\left.\left.+4 \text{e}^{\frac{1024 \alpha ^8+96 \pi ^2 \alpha ^4+\pi ^4}{256 \alpha ^6+16 \pi ^2
   \alpha ^2}}-2 \sqrt{3}\right)\right)\right]\right\},
\end{aligned}   
\end{equation}

\begin{equation}
\begin{aligned}
&\langle\hat\chi_{\KS}(1,1/2)\rangle=\frac{
\text{e}^{-\frac{768 \alpha ^8+96 \pi ^2 \alpha ^4+\pi ^4}{128 \alpha ^6+8 \pi ^2 \alpha ^2}}}{2 \sqrt{\text{e}^{-4 \alpha ^2}+1}}
   \left\{2 \sqrt{2} \left(\text{e}^{\frac{2 \alpha ^2 \left(32 \alpha ^4+\pi ^2\right)}{16 \alpha ^4+\pi
   ^2}}+1\right) \text{e}^{\frac{1024 \alpha ^8+320 \pi ^2 \alpha ^4+3 \pi ^4}{512 \alpha ^6+32
   \pi ^2 \alpha ^2}} \cos \left(\frac{4 \pi  \alpha ^4}{16 \alpha ^4+\pi
   ^2}\right)\right.\\&\left.+\text{e}^{\frac{512 \alpha ^8+\pi ^4}{256 \alpha ^6+16 \pi ^2 \alpha ^2}} \left[2 \sqrt{2}
   \text{e}^{\frac{320 \pi ^2 \alpha ^4+\pi ^4}{512 \alpha ^6+32 \pi ^2 \alpha ^2}} \left(\text{e}^{\frac{2
   \alpha ^2 \left(32 \alpha ^4+\pi ^2\right)}{16 \alpha ^4+\pi ^2}}+1\right) \sin \left(\frac{4
   \pi  \alpha ^4}{16 \alpha ^4+\pi ^2}\right)+\left(\text{e}^{\frac{12 \pi ^2 \alpha ^2}{16 \alpha
   ^4+\pi ^2}}+\text{e}^{\frac{8 \alpha ^2 \left(8 \alpha ^4+\pi ^2\right)}{16 \alpha ^4+\pi
   ^2}}{+}\text{e}^{\frac{64 \alpha ^6+12 \pi ^2 \alpha ^2}{16 \alpha ^4+\pi ^2}}{+}1\right)\right.\right.\\&\left.\left.\times \sin
   \left(\frac{8 \pi  \alpha ^4}{16 \alpha ^4+\pi ^2}\right)+2 \text{e}^{\frac{6 \pi ^2 \alpha ^2}{16
   \alpha ^4+\pi ^2}} \left(\text{e}^{4 \alpha ^2}+\text{e}^{\frac{6 \pi ^2 \alpha ^2}{16 \alpha ^4+\pi
   ^2}}+\text{e}^{\frac{64 \alpha ^6+6 \pi ^2 \alpha ^2}{16 \alpha ^4+\pi ^2}}+\text{e}^{\frac{32 \pi ^2 \alpha
   ^4+\pi ^4}{256 \alpha ^6+16 \pi ^2 \alpha ^2}}+\text{e}^{\frac{1024 \alpha ^8+96 \pi ^2 \alpha ^4+\pi
   ^4}{256 \alpha ^6+16 \pi ^2 \alpha ^2}}-1\right)\right]\right\},
\end{aligned}   
\end{equation}

\begin{equation}
\begin{aligned}
&\langle\hat\chi_{\KS}(1/2,1/2)\rangle=\frac{\text{e}^{-\frac{1024 \alpha ^8+192 \pi ^2 \alpha ^4+\pi ^4}{16 \alpha ^6+\pi ^2 \alpha ^2}}}{8 \sqrt{\text{e}^{-4 |\alpha |^2}+1}}
   \left(\text{e}^{\frac{1024 \alpha ^8+192 \pi ^2 \alpha ^4+\pi ^4}{16 \alpha ^6+\pi ^2 \alpha
   ^2}}\right)^{15/16} \left[\text{e}^{\frac{8 \alpha ^2 \left(8 \alpha ^4+\pi ^2\right)}{16 \alpha
   ^4+\pi ^2}} \sqrt{\text{e}^{-4 |\alpha |^2}+1}+2 \text{e}^{\frac{64 \alpha ^6+10 \pi ^2 \alpha ^2}{16 \alpha
   ^4+\pi ^2}} \sqrt{\text{e}^{-4 |\alpha |^2}+1}\right.\\&\left.+3 \text{e}^{\frac{64 \alpha ^6+12 \pi ^2 \alpha ^2}{16 \alpha
   ^4+\pi ^2}} \sqrt{\text{e}^{-4 |\alpha |^2}+1}+\left(\text{e}^{\frac{8 \alpha ^2 \left(8 \alpha ^4+\pi
   ^2\right)}{16 \alpha ^4+\pi ^2}} \left(\sqrt{\text{e}^{-4 |\alpha |^2}+1}+2\right)+\text{e}^{\frac{64 \alpha
   ^6+12 \pi ^2 \alpha ^2}{16 \alpha ^4+\pi ^2}} \left(3 \sqrt{\text{e}^{-4 |\alpha |^2}+1}+2\right)\right.\right.\\&\left.\left.+2
   \text{e}^{\frac{64 \alpha ^6+10 \pi ^2 \alpha ^2}{16 \alpha ^4+\pi ^2}} \sqrt{\text{e}^{-4 |\alpha |^2}+1}+2
   \text{e}^{\frac{12 \pi ^2 \alpha ^2}{16 \alpha ^4+\pi ^2}}+2\right) \sin \left(\frac{8 \pi  \alpha
   ^4}{16 \alpha ^4+\pi ^2}\right)+8 \text{e}^{\frac{320 \pi ^2 \alpha ^4+\pi ^4}{512 \alpha ^6+32 \pi
   ^2 \alpha ^2}} \left(\text{e}^{\frac{2 \alpha ^2 \left(32 \alpha ^4+\pi ^2\right)}{16 \alpha ^4+\pi
   ^2}} \left(\sqrt{\text{e}^{-4 |\alpha |^2}+1}+1\right)+1\right)\right.\\&\left. \cos \left(\frac{\pi ^3}{64 \alpha
   ^4+4 \pi ^2}\right)+4 \text{e}^{\frac{1024 \alpha ^8+192 \pi ^2 \alpha ^4+\pi ^4}{256 \alpha ^6+16
   \pi ^2 \alpha ^2}} \sqrt{\text{e}^{-4 |\alpha |^2}+1}-4 \text{e}^{\frac{6 \pi ^2 \alpha ^2}{16 \alpha ^4+\pi
   ^2}}+4 \text{e}^{\frac{12 \pi ^2 \alpha ^2}{16 \alpha ^4+\pi ^2}}+4 \text{e}^{\frac{64 \alpha ^6+10 \pi ^2
   \alpha ^2}{16 \alpha ^4+\pi ^2}}\right.\\&\left.+4 \text{e}^{\frac{64 \alpha ^6+12 \pi ^2 \alpha ^2}{16 \alpha ^4+\pi
   ^2}}+4 \text{e}^{\frac{128 \pi ^2 \alpha ^4+\pi ^4}{256 \alpha ^6+16 \pi ^2 \alpha ^2}}+4
   \text{e}^{\frac{1024 \alpha ^8+192 \pi ^2 \alpha ^4+\pi ^4}{256 \alpha ^6+16 \pi ^2 \alpha
   ^2}}\right],
\end{aligned}   
\end{equation}

\begin{equation}
\begin{aligned}
&\langle\hat\chi_{\KS}(0,1/2)\rangle=
\frac{1}{4} \text{e}^{-\frac{64 \pi ^2 \alpha ^4+\pi ^4}{256 \alpha ^6+16 \pi ^2 \alpha ^2}} \left[2
   \text{e}^{\frac{2 \pi ^2 \alpha ^2}{16 \alpha ^4+\pi ^2}}+3 \text{e}^{\frac{4 \pi ^2 \alpha ^2}{16 \alpha
   ^4+\pi ^2}}+\left(2 \text{e}^{\frac{2 \pi ^2 \alpha ^2}{16 \alpha ^4+\pi ^2}}+3 \text{e}^{\frac{4 \pi ^2
   \alpha ^2}{16 \alpha ^4+\pi ^2}}+1\right) \sin \left(\frac{8 \pi  \alpha ^4}{16 \alpha ^4+\pi
   ^2}\right)\right.\\&\left.+4 \text{e}^{\frac{64 \pi ^2 \alpha ^4+\pi ^4}{256 \alpha ^6+16 \pi ^2 \alpha ^2}}+8
   \text{e}^{\frac{128 \pi ^2 \alpha ^4+\pi ^4}{512 \alpha ^6+32 \pi ^2 \alpha ^2}} \cos \left(\frac{\pi
   ^3}{64 \alpha ^4+4 \pi ^2}\right)+1\right].
\end{aligned}   
\end{equation}
\end{widetext}

These functions are used in order to produce the plots shown in Fig.~\ref{wernerfig} {\bf (a)} and {\bf (b)}.

%%%

%%%%%


\begin{thebibliography}{99}
\bibitem{Specker1} E. P. Specker, Dialectica {\bf 14}, 239 (1960).
\bibitem{Specker2} S. Kochen and E. P. Specker, J. Math. Mech. {\bf 17},  59
(1967).
\bibitem{Specker3} J. S. Bell, Rev. Mod. Phys. {\bf 38}, 447 (1966).
\bibitem{bell} J. S. Bell, Physics {\bf 1}, 195 (1964); J. S. Bell, {\it Speakable and Unspeakable in Quantum Mechanics} (Cambridge University Press 1987) 
\bibitem{peres} A. Peres, Phys. Lett. A {\bf 151}, 107 (1990).
\bibitem{mermin} N. D. Mermin, Phys. Rev. Lett. {\bf 65}, 3373 (1990).
\bibitem{guhene} O. G\"uhne, M. Kleinmann, A. Cabello, J.-A. Larsson, G. Kirchmair, F. Z\"ahringer, R. Gerritsma, and C. F. Roos, Phys. Rev. A {\bf 81}, 022121 (2010).
\bibitem{cabello} A. Cabello, Phys. Rev. Lett. {\bf 101}, 210401 (2008).
\bibitem{Kirchmair} G. Kirchmair, F. Z\"{a}hringer, R. Gerritsma, M. Kleinmann, O. G\"{u}hne, A. Cabello, R. Blatt, and C. F. Roos, Nature {\bf 460}, 494 (2009).
%\bibitem{papers cited by state indep quantum context for cv variables} 
%{\bf what do you mean? 2nd paragrah introduction}.
\bibitem{Plastino} \'{A}. R. Plastino and A. Cabello, Phys. Rev. A {\bf 82}, 022114 (2010).
\bibitem{Banaszek} K. Banaszek and K. W\'{o}dkiewicz, Phys. Rev. A {\bf 58}, 4345 (1998); Phys. Rev. Lett. {\bf 82}, 2009 (1999).
\bibitem{Chen} Z. B. Chen, J. W. Pan, G. Hou, and Y. D. Zhang, Phys. Rev. Lett. {\bf 88}, 040406 (2002).
\bibitem{Jeong} H. Jeong, W. Son, M. S. Kim, D. Ahn, and C. Brukner, Phys. Rev. A {\bf 67}, 012106 (2003).
\bibitem{Paternostro} M. Paternostro, H. Jeong, and T. C. Ralph, Phys. Rev. A {\bf 79}, 012101 (2009).
\bibitem{sanders} B. C. Sanders, Phys. Rev. A {\bf 45}, 6811 (1992); B. C. Sanders, K. S. Lee, and M. S. Kim, Phys. Rev. A {\bf 52}, 735 (1995).
\bibitem{Stobinska} M. Stobi\'nska, H. Jeong, and T. C. Ralph, Phys. Rev. A {\bf 75}, 052105 (2007); H. Jeong and Nguyen Ba An, {\it ibid.} {\bf 74}, 022104 (2006).
\bibitem{jeong} H. Jeong, M. Paternostro, and T. C. Ralph, Phys. Rev. Lett. {\bf 102}, 060403 (2009).
\bibitem{MP} M. Paternostro and H. Jeong, Phys. Rev. A {\bf 81}, 032115 (2010).
\bibitem{lee} C.-W. Lee, M. Paternostro, and H. Jeong, Phys. Rev. A {\bf 83}, 022102 (2011).
\bibitem{GMcK} G. McKeown, F. L. Semi\~ao, H. Jeong, and M. Paternostro, Phys. Rev. A {\bf 82}, 022315 (2010).
\bibitem{ban02} K. Banaszek, A. Dragan, K. W\'odkiewicz, and C. Radzewicz,
Phys. Rev. A {\bf 66}, 043803 (2002).
\bibitem{cinv05} C. Invernizzi, S. Olivares, M. G. A. Paris, and K. Banaszek, 
Phys. Rev. A {\bf 72}, 042105 (2005).
\bibitem{so05} S. Olivares, and M. G. A. Paris, J. Opt. B: Quantum Semiclass. 
Opt. {\bf 7}, 392 (2005). 
%\bibitem{Janszky} J. Janszky and An. V. Vinogradov, Phys. Rev. Lett. {\bf 64}, 2771 (1990); V. Bu\v{z}ek, A. Vidiella-Barranco, and P. L. Knight, Phys. Rev. A {\bf 45}, 6570 (1992).
\bibitem{Glauber}  R. J. Glauber, Phys. Rev. {\bf 131}, 2766 (1963).
\bibitem{barnett} S. M. Barnett and P. M. Radmore,
 {\it Methods in Theoretical Quantum Optics} (Clarendon, Oxford, 1997).
\bibitem{walls} D. F. Walls and G. J. Milburn, {\it Quantum Optics} (Springer, Heidelberg, 2008).
\bibitem{knight} J. Janszky and A. V. Vinogradov, Phys. Rev. Lett. {\bf 64}, 2771 (1990); V. Bu\v{z}ek, A. Vidiella-Barranco, and P. L. Knight, Phys. Rev. A {\bf 45}, 6570 (1992).
\bibitem{Fer05} A. Ferraro, and M. G. A. Paris,  J. Opt. B: Quantum Semiclass. Opt. {\bf 7}, 
174 (2005).
\bibitem{gis92} N. Gisin, and A. Peres, Phys. Lett. A {\bf 162}, 15 (1992).
\bibitem{incorso} G. McKeown {\it et al.} (to be published).
\bibitem{blencowe} M. P. Blencowe, Contemp. Phys. {\bf 46}, 249 (2005).
\bibitem{engineering} K. Jacobs, L. Tian, and J. Finn, Phys. Rev. Lett. {\bf 102}, 057208 (2009); K. Jacobs and A. Landahl, Phys. Rev. Lett. {\bf 103}, 067201 (2009); F. L. Semi\~ao, K. Furuya, and G. J. Milburn, Phys. Rev. A {\bf 79}, 063811 (2009).
\bibitem{jacobs} F. W. Strauch, K. Jacobs, and R. W. Simmonds, Phys. Rev. Lett. {\bf 105}, 050501 (2010); F. Marquardt, Phys. Rev. B {\bf 76}, 205416 (2007).
\end{thebibliography}
\end{document}